\documentclass[twocolumn,superscriptaddress,nofootinbib,notitlepage,longbibliography,aps]{revtex4-2}
\pdfoutput=1

\usepackage{graphicx}
\usepackage{amsmath}
\usepackage{amssymb}
\usepackage{amsfonts}
\usepackage[dvipsnames]{xcolor}
\usepackage[linktoc=none]{hyperref}
\usepackage[utf8]{inputenc}
\usepackage{booktabs}
\usepackage{comment}
\usepackage{multirow}
\usepackage{float}
\definecolor{tab-blue}{RGB}{0, 107, 164}
\hypersetup{colorlinks=true,allcolors=tab-blue}

\newcommand{\INFN}{INFN - Sezione di Napoli, Complesso Universitario Monte Sant'Angelo, 80126 Napoli, Italy}
\newcommand{\SSM}{Scuola Superiore Meridionale, Via Mezzocannone 4, 80138 Napoli, Italy}

\begin{document}
\title{Asteroid-mass Primordial Black Holes as Dark Matter from Supersymmetry}

\author{Andrea Boccia}
\email{andrea.boccia-ssm@unina.it}
\affiliation{\SSM}
\affiliation{\INFN}
\author{Marco Chianese}
\email{m.chianese@ssmeridionale.it}
\affiliation{\SSM}
\affiliation{\INFN}

\begin{abstract}
We study the formation of asteroid-mass Primordial Black Holes (PBHs) as a dark matter candidate in supersymmetric extensions of the Standard Model. We show that the presence of heavy particles predicted in the Minimal Supersymmetric Standard Model (MSSM) can lead to a transient softening of the equation of state of the Universe during their non-relativistic transition, enhancing PBH formation. We compute the effective equation of state for different realizations of the MSSM mass spectrum, parametrized by three characteristic mass scales. Assuming a broad and approximately scale-invariant primordial curvature power spectrum, we evaluate the resulting PBH mass functions and compare them with current observational constraints. We find that, for supersymmetric masses above $\sim 10^5\,\mathrm{GeV}$, the PBH mass function is significantly enhanced in the asteroid-mass window, allowing PBHs to account for the total dark matter abundance without violating existing bounds. In contrast, within the Standard Model the same configurations lead to PBH mass functions that are observationally excluded. For lighter supersymmetric mass spectra, PBH production is shifted toward masses above $\sim 10^{22}\,\mathrm{g}$, which are strongly constrained by microlensing searches, thereby reducing their  allowed contribution to the dark matter density.
\end{abstract}
\maketitle

\section{Introduction}
\label{sec:intro}
Multiple independent cosmological and astrophysical observations provide compelling evidence for a cold, non-baryonic component of the Universe, commonly referred to as Dark Matter (DM). Yet, despite decades of theoretical and experimental efforts, its fundamental nature remains unknown and constitutes one of the most pressing open questions in modern physics. A wide range of well-motivated particle candidates, like weakly interacting massive particles~\cite{Jungman:1995df, Bertone:2004pz} and axion-like particles~\cite{Marsh:2015xka, Choi:2020rgn}, has been proposed, however, no conclusive evidence has emerged from collider searches, direct detection experiments, or indirect probes. This situation motivates the investigation of alternative, non-particle explanations for dark matter. Primordial Black Holes (PBHs) emerge as a particularly well-motivated alternative.

First proposed by Zel’dovich \cite{Zeldovich:1967lct}, Hawking \cite{Hawking:1971ei} and Carr \cite{Carr:1974nx}, PBHs can form in the early Universe, shortly after inflation, {\it e.g.} through the gravitational collapse of large primordial density fluctuations. Unlike astrophysical black holes, whose origin is tied to stellar evolution, PBHs would arise from purely cosmological processes and could therefore span an extremely wide mass range, determined by the horizon mass at the time of formation. If sufficiently massive to survive Hawking evaporation \cite{Hawking:1975vcx} until the present epoch, they would behave as a cold, non-baryonic component of the matter density, thus constituting a viable DM candidate~\cite{Carr:2016drx, Green:2020jor}. 
Observationally, black holes are known to exist over a broad range of masses. Stellar-mass black holes are well established as the end products of massive star evolution, while supermassive black holes, with masses ranging from millions to billions of solar masses, are known to reside at the centers of most galaxies. More recently, gravitational-wave observations by the LIGO/Virgo collaborations \cite{LIGOScientific:2016sjg} have revealed merging black holes with masses $M \gtrsim 30 \, M_\odot$, further stimulating interest in the possible cosmological origin of some of these objects \cite{Bird:2016dcv, Garcia-Bellido:2017fdg, Sasaki:2018dmp, Bartolo:2018evs}. Over the years, an extensive body of work on PBH phenomenology has placed increasingly stringent constraints on their cosmological abundance across a wide range of observational probes. Current bounds rule out PBHs as the dominant DM component over most of the parameter space. However, a viable window remains in the asteroid-to-planetary mass range, $10^{18}\,\mathrm{g} \lesssim M_{\rm PBH} \lesssim 10^{22}\,\mathrm{g}$. This region is bounded at lower masses by evaporation constraints \cite{Lehmann:2018ejc, Laha:2019ssq, Boudaud:2018hqb, Clark:2016nst, Dasgupta:2019cae, Laha:2020ivk, Coogan:2020tuf, Kim:2020ngi, Calabrese:2021zfq, Mittal:2021egv, Su:2024hrp, Tan:2024nbx, DelaTorreLuque:2024qms, Khan:2025kag} and at higher masses by microlensing observations \cite{Macho:2000nvd, EROS-2:2006ryy, Garcia-Bellido:2017imq, Oguri:2017ock, Croon:2020ouk, Blaineau:2022nhy, Leung:2022vcx, Esteban-Gutierrez:2023qcz, Mroz:2024wia}. Recent studies have pointed out that quantum effects, referred to as ``memory burden''~\cite{Dvali:2018xpy, Dvali:2020wft}, may slow down black hole evaporation, allowing much lighter PBHs to survive until today and account for the totality of the dark matter~\cite{Alexandre:2024nuo, Barman:2024slw, Thoss:2024hsr, Chianese:2024rsn, Zantedeschi:2024ram, Boccia:2025hpm, Calabrese:2025sfh, Chianese:2025wrk, Ambrosone:2026djo}. We do not consider this scenario here.

If PBHs form through gravitational collapse in the early Universe, their abundance is primarily controlled by two key ingredients: the primordial curvature power spectrum $\mathcal{P}_\zeta$ and the equation of state (EoS) of the primordial plasma at the time of horizon re-entry. The former is dictated by the inflationary dynamics and may exhibit enhanced features on the small scales relevant for PBH formation. The latter, instead, is determined by the thermal history of the Universe and ultimately by its underlying particle content.
In particular, cosmological phase transitions or the decoupling of particle species from the thermal bath can induce a temporary softening of the EoS from its value $w = p/\rho =  1/3$ in radiation domination, lowering the collapse threshold and exponentially enhancing the PBH production at masses corresponding to the horizon mass at that epoch. Several works~\cite{Jedamzik:1996mr, Jedamzik:1998hc, Widerin:1998my, Byrnes:2018clq, Carr:2019hud, Conzinu:2020cke, Conzinu:2023fui,Musco:2023dak, Gonin:2026xhe} have shown that the EoS softening during the QCD phase transition can enhance near-solar mass PBH formation by orders of magnitude, given sufficient power spectrum amplitude at the relevant scales.
Similar effects on different mass ranges can be achieved by considering the softening of the EoS due to electroweak phase transition or electron-positron annihilation \cite{Pritchard:2024vix,Blas:2026xws}.
Moreover, several studies~\cite{Allahverdi:2020bys, Harada:2016mhb, Harada:2017fjm, Kokubu:2018fxy, Harada:2022xjp} have also explored scenarios in which PBHs form during an early matter-dominated phase, characterized by an EoS parameter $w \simeq 0$, leading to a significant enhancement of PBH production by several orders of magnitude. 

In this work, we investigate the impact on the equation of state induced by a particle sector characterized by a large number of degrees of freedom, becoming non-relativistic at temperatures above the electroweak scale, as motivated by beyond-the-Standard-Model scenarios. In particular, we focus on the Minimal Supersymmetric Standard Model (MSSM), which predicts a superpartner for each Standard Model (SM) particle \cite{Haber:1993wf, Baer:1995tb, Drees:1995hj, Csaki:1996ks}. We show that the resulting temporary softening of the equation of state at temperatures $T \gtrsim 1\,\mathrm{TeV}$ can significantly enhance the PBH abundance at formation within a mass range compatible with the so-called asteroid-mass window. Similar modifications of the thermal history above the electroweak scale have been considered in previous studies, along with their implications for the PBH formation \cite{Escriva:2023nzn, Pritchard:2025pcn} and for the production of gravitational waves from cosmic string networks~\cite{Cui:2018rwi, Antusch:2024nqg}.

We here adopt a self-consistent treatment of PBH formation, explicitly incorporating the dependence of the collapse threshold on the equation of state across the relevant temperature range and analyzing the resulting extended mass functions in light of current observational constraints. Differently from Ref.~\cite{Pritchard:2025pcn}, which focuses on a single supersymmetric breaking scale, we consider a more general MSSM setup characterized by three independent mass scales: a single mass scale $m_f$ governing the fermionic superpartners, a single mass scale $m_b$ governing the bosonic superpartners, and a separate scale $m_h$ for the second Higgs doublet required in the MSSM. This multi-scale approach enables a systematic exploration of the parameter space and allows us to identify the regions where departures from the SM thermal history can lead to PBH mass functions that remain compatible with existing bounds.

The paper is organized as follows. In Sec.~\ref{sec:eos}, we outline the computation of the EoS parameter for a general particle content and introduce the MSSM setup considered in this work. In Sec.~\ref{sec:formation}, we review the basics of the PBH formation via critical collapse and describe the procedure adopted to confront extended mass functions with the monochromatic observational constraints. In Sec.~\ref{sec:res}, we present our results and compare them with the ones obtained within the SM. Finally, in Sec.~\ref{sec:conc}, we summarize our findings and discuss their implications.

%====================================
\section{Universe's equation of state within the MSSM}
\label{sec:eos}
%====================================

The thermal history of the Universe is fully determined by the particle content in equilibrium with the primordial plasma. Following the reheating epoch, the Universe is into a state characterized by a hot, thermalized plasma composed of ultra-relativistic species. In this regime, both energy density $\rho(T)$ and pressure $p(T)$ evolve according to the radiation-dominated scaling, and the equation of the state of the Universe is equal to the one of a relativistic fluid. This behavior persists until the cosmic temperature falls below the rest-mass threshold of individual particle species, at which point each species undergoes a relativistic-to-non-relativistic transition. This induces changes in the equation of state of the plasma, leaving possible observable imprints on several cosmological processes. 

Let us first describe the general methodology to compute the equation of state as a function of the temperature $T$ of the thermal bath. The EoS parameter can be written as 
\begin{equation}
\label{eq:eos}
    w(T) = \frac{p(T)}{\rho(T)} = \frac{4}{3} \frac{g_s(T)}{g_*(T)} - 1\,, 
\end{equation}
where $ g_s(T)$ and $g_*(T)$ are the effective degrees of freedom associated to entropy and energy density, respectively. The effective degrees of freedom control the expansion rate through the Friedmann equation and determine the relation between temperature and cosmic time. When all particle species are ultra-relativistic and in thermal equilibrium, one has $g_s(T)=g_*(T)$ and therefore $w = 1/3$. Deviations occur when species become non-relativistic or when different sectors have different temperatures, as during annihilations, decouplings, or phase transitions.
In the standard cosmological scenario, this occurs several times during the history of the Universe according to the SM particle content. The biggest drop in $w$ is expected during the QCD phase transition due to the entropy being redistributed among the high number of hadrons that become available when the temperature of the thermal bath drops below $T_{\rm QCD} \sim 200 \, \rm MeV$~\cite{Borsanyi:2013bia, HotQCD:2014kol}.

The effective degrees of freedom can be computed from the energy, pressure, and entropy densities of a relativistic fluid composed of multiple particle species with masses $m_j$ in thermal equilibrium at temperature $T$. Introducing the auxiliary variables $u = E/T$ and $z_j = m_j/T$, one obtains the corresponding expressions for the effective degrees of freedom associated with energy, pressure, and entropy~\cite{Husdal:2016haj}
\begin{equation}
\label{eq:dof}
    \begin{split}
        g_*(T) & = \sum_j \frac{15}{\pi^4}g_j \int_{z_j}^\infty \frac{u^2\sqrt{u^2 - z_j^2}}{e^u \pm 1} {\rm d} u\,,\\
        g_p(T) &= \sum_j \frac{15}{\pi^4}g_j \int_{z_j}^\infty \frac{(u^2 - z_j^2)^{3/2}}{e^u \pm 1} {\rm d} u\,,\\
        g_s(T) &= \frac{3g_*(T) + g_p(T)}{4}\,,
    \end{split}
\end{equation}
where $g_j$ are the internal degrees of freedom of each particle species and the $\pm$ sign in the denominator accounts for Pauli blocking (plus sign for fermions) and Bose enhancement (minus sign for bosons), with the sum being performed over all particle species. The integrals in Eqs.~\eqref{eq:dof} can be evaluated numerically as a function of temperature for each
particle species.
Then, Eq.~\eqref{eq:eos} can be used to obtain the EoS parameter for any beyond-the-Standard-Model scenario.
\begin{figure}[t!]
    \centering
    \includegraphics[width=0.95\linewidth]{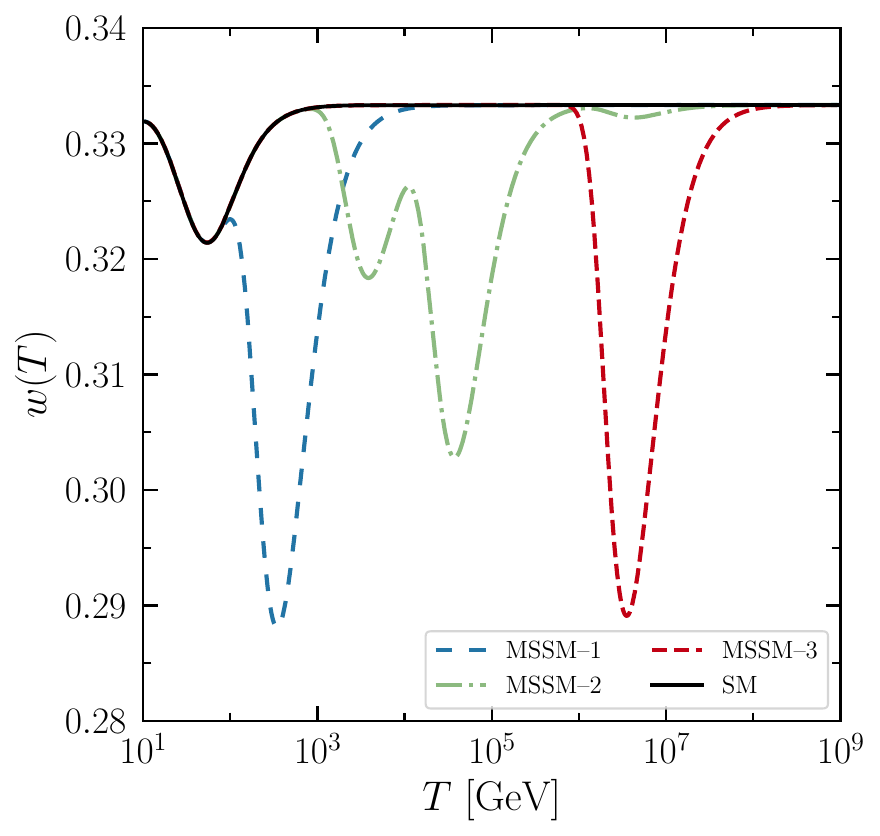}
    \caption{Equation of state parameter as a function of temperature for three supersymmetric scenarios: a degenerate mass spectrum at low (long-dashed blue line) and high (short-dashed red line) scales, and a non-degenerate spectrum with intermediate mass splitting (dot-dashed green line). Also shown is the equation of state for the SM scenario (solid black line).}
    \label{fig:w}
\end{figure}

In the MSSM, each SM particle has a superpartner with the same quantum numbers but with spin differing by one-half. Moreover, a second Higgs doublet is introduced because the superpotential must be holomorphic, preventing the use of the conjugate Higgs field to generate masses for both up- and down-type quarks. A second Higgs doublet is also required to cancel gauge anomalies introduced by the Higgsino fermions. Hence, at very high temperatures, when all particles are in thermal equilibrium, the MSSM has $g_* = 228.75$ relativistic degrees of freedom, {\it i.e.} nearly twice the SM value of $g_* = 106.75$. In general, it is assumed that all the supersymmetric particles share a common mass scale. This is however a simplifying assumption, as the Soft Supersymmetric Breaking (SSB) parameters responsible for breaking supersymmetry down to the SM generically arise from different terms in the Lagrangian, naturally leading to a non-degenerate mass spectrum. These soft parameters can be divided into three classes (see {\it e.g.} Ref.~\cite{Berezhiani:2015vea}): the $F$-terms, which give masses of the order $\mathcal{O}(M_F)$ to the fermions including the Majorana gaugino masses; the $D$-terms, which give masses of the order $\mathcal{O}(M_D)$ to the bosons including the squarks, the sleptons and the Higgses; and the so-called $\mu$-term, which determines the Higgsino masses and contributes to the masses of the two Higgs doublets. Embedding the MSSM in the framework of a Grand Unified Theory (GUT) imposes additional constraints on the MSSM mass spectrum~\cite{Dimopoulos:1981yj, Ibanez:1981yh, Einhorn:1981sx, Marciano:1981un, Berezhiani:2015vea}.  Indeed, GUT unification requires all gaugino masses to be equal at the GUT scale, and an analogous mass unification can be assumed for squarks and sleptons belonging to the same GUT multiplet. The mass differences observed among gluinos and neutralinos, or among squarks and sleptons, therefore arise purely from renormalization group running, which generically leads to order-one modifications of the MSSM masses~\cite{Castano:1993ri, Berezhiani:2015vea, Pritchard:2025pcn}.

In the present work, rather than focusing on a specific realization of the MSSM, we consider a phenomenological scenario characterized by three distinct mass scales for the MSSM particles: a common mass scale $m_f \sim \mathcal{O}(M_F)$ for the fermionic superpartners, a common mass scale $m_b \sim \mathcal{O}(M_D)$ for the bosonic superpartners, and a mass scale $m_h$ for the second decoupled Higgs doublet. The mass scale $m_h$ indeed arises from a combination of all the three SSB parameters, which together define the full mass matrix of the two Higgs doublets. This simplified framework, which emerges in some concrete supersymmetric realizations (see {\it e.g.} Refs.~\cite{Giudice:2004tc, Arkani-Hamed:2004zhs}), allows us to perform a systematic exploration of the modifications to the EoS parameter and, consequently, to the PBH mass function arising from a non-degenerate MSSM mass spectrum.

In Fig.~\ref{fig:w} we show the EoS parameter as a function of temperature for the following different MSSM benchmark realizations: 
\begin{itemize}
    \item {\bf low-scale degenerate spectrum (MSSM--1)} with $m_f = m_b = m_h = 10^3~{\rm GeV}$;
    \item {\bf hierarchical spectrum (MSSM--2)} with $m_f = 10^4~{\rm GeV}$, $m_b = 10^5~{\rm GeV}$, and $m_h = 10^7~{\rm GeV}$;
    \item {\bf high-scale degenerate spectrum (MSSM--3)} with $m_f = m_b = m_h = 10^7~{\rm GeV}$.
\end{itemize}
We find that the most significant deviation from the radiation-dominated value $w = 1/3$ occurs in the case of degenerate MSSM masses, where $w\simeq 0.289$. In contrast, a non-degenerate spectrum leads to a series of smaller departures from $w = 1/3$, each associated with a distinct feature in the PBH mass function, as will be discussed in the following sections. In this scenario, the largest deviation arises around $T \sim m_b$, corresponding to the mass scale at which the number of additional MSSM degrees of freedom is maximized. Quantitatively, we observe a reduction of approximately $\sim 10\%$, $\sim 5\%$, and $\lesssim 1\%$ at $m_b$, $m_f$, and $m_h$, respectively.
We also show the EoS parameter for the SM scenario (black solid line), where the electroweak sector induces $w < 1/3 $ at temperatures $T \simeq 100~{\rm GeV}$.

%====================================
\section{Primordial black holes formation and mass function}
\label{sec:formation}
%====================================

PBH formation has been widely discussed and many mechanisms have been proposed throughout the years~\cite{Flores:2024eyy}. Here we briefly review the formation through direct gravitational collapse of density anisotropies in the early universe within the threshold statistics formalism~\cite{Young:2019yug,Franciolini:2023pbf,Gow:2022jfb}.

Given the primordial curvature power spectrum $\mathcal{P}_\zeta(k)$ resulting from inflation, the corresponding power spectrum of the primordial density contrast, $\delta = \delta \rho / \rho$, can be derived as~\cite{Green:2004wb}
\begin{equation}
\mathcal{P}_{\delta}(k ) = \left(\frac{4}{9}\right)^2 (k R)^4 \tilde{W}^2(k R) T^2(k R) \mathcal{P}_\zeta(k)\,,
\end{equation}
where
\begin{equation}
\tilde{W}(k R) = 3 \frac{\sin(k R) - (k R)\cos(k R)}{(k R)^3}
\end{equation}
is the Fourier transform of the real-space top-hat window function used to smooth the density contrast over the comoving scale $R = r_m/(aH)$, with $r_m = 3.4$ being a phenomenological parameter fixed following Refs.~\cite{Musco:2020jjb, Musco:2021sva, Byrnes:2018clq, Ianniccari:2024ltb}, and
\begin{equation}
T(k R) = 3 \frac{\sin(k R/\sqrt{3}) - (k R/\sqrt{3})\cos(k R/\sqrt{3})}{(k R/\sqrt{3})^3}
\end{equation}
is the radiation transfer function~\cite{Josan:2009qn}.
When the density perturbations $\delta$ re-enter the horizon after inflation, they may collapse to form a black hole if their amplitude exceeds a critical threshold $\delta_c$. This threshold depends on the details of the collapse, including the shape of the perturbation and the equation of state of the background fluid, and is typically determined through numerical hydrodynamical simulations~\cite{Musco:2004ak, Musco:2012au, Musco:2018rwt, Musco:2020jjb, Musco:2021sva, Escriva:2024aeo}. If the amplitude of the primordial perturbation $\delta$ is close to the threshold value, the relation between the PBH mass and the horizon mass at the time of formation is given by \cite{Choptuik:1992jv,Niemeyer:1997mt,Musco:2004ak}
\begin{equation}
    \label{eq:masses}
    M_{\rm PBH} (T)= \kappa (\delta - \delta_c)^\gamma M_{\rm H}(T)\,,
\end{equation}
where $\kappa = 3.3$ and $\gamma = 0.36$ are numerically estimated parameters encoding the details of the collapse. Here, we assume that only the collapse threshold $\delta_c (\omega(T))$ depends on the EoS parameter, taking the numerical results for $\delta_c (\omega(T))$ from Ref.~\cite{Musco:2012au}. The relation between the horizon mass $M_{\rm H}$ and the temperature of the thermal bath in a radiation dominated universe is ~\cite{Wang:2019kaf,Nakama:2016gzw, Carr:2023tpt} 
\begin{equation}
    \label{eq:temp}
    M_{\rm  H}(T) = 12 \, \left(\frac{10}{g_*(T)}\right)^{1/2} \frac{M_P}{T^2}\,,
\end{equation}
with $M_P$ being the reduced Planck mass. 
Relying on the one-to-one correspondence given by Eq.~\eqref{eq:temp}, throughout this section we will use interchangeably the horizon mass $M_H$ and the corresponding temperature $T_H \equiv T(M_H)$.
Using the relation between the wavenumber $k$ at horizon re-entry and the horizon mass~\cite{Wang:2019kaf}
\begin{equation}
\label{eq:pivot}
\frac{k}{k^*} \simeq 83.5 \left(\frac{g_*(T_{\rm H})}{106.75}\right)^{ \frac14}\left(\frac{g_s(T_{\rm H})}{106.75}\right)^{ \frac13}\left(\frac{10^{20} \rm g}{M_{\rm H}}\right)^{\frac12},
\end{equation}
with $k^* = 10^{10}~\rm Mpc^{-1}$, we can directly compute the fraction of dark matter in the form of PBHs of mass $M_{\rm PBH}$.

Assuming Gaussian statistics, the probability distribution of the density contrast is given by
\begin{equation}
P(\delta) = \frac{1}{\sqrt{2\pi\sigma^2(M_{\rm H})}}
\exp\left(-\frac{\delta^2}{2 \sigma^2(M_{\rm H})}\right) ,
\end{equation}
where the variance $\sigma^2(M_{\rm H})$ is obtained from the power spectrum as~\cite{Young:2014ana}
\begin{equation}
\sigma^2(M_{\rm H}) = \int_0^{\infty} \mathcal{P}_{\delta}(k ) \mathrm{d}\ln k \, .
\end{equation}
Primordial non-Gaussianities may also be considered and can significantly affect the resulting PBH mass function~\cite{Bullock:1996at,Ivanov:1997ia,PinaAvelino:2005rm,Byrnes:2012yx,Shandera:2012ke,Young:2013oia,Young:2015cyn,Franciolini:2023pbf}. While we do not include such primordial non-Gaussianities in this work, we do account for the unavoidable non-linear effects discussed in~\cite{DeLuca:2019qsy,Allegrini:2024ooy}.
Following~\cite{Allegrini:2024ooy}, we compute the fraction of dark matter in the form of PBHs of mass $M_{\rm PBH}$ as
\begin{equation}
\label{eq:fofm}
\begin{aligned}
\mathcal{F}(M_{\rm PBH}) &= \frac{1}{\Omega_{\rm DM}}
\int_{M_{\rm min}}^{\infty} \frac{\mathrm{d} M_{\rm H}}{M_{\rm H}}\,
\sqrt{\frac{M_{\rm eq}}{M_{\rm H}}}\,\frac{\kappa}{\gamma}\, \mu^{\frac{1+\gamma}{\gamma}}\\
&\times \frac{1}{\sqrt{2\pi\sigma^2 \Lambda}}\exp\left[-\frac{8\left(1- \sqrt{\Lambda})^2\right)^2}{9 \sigma^2}\right] \,,
\end{aligned}
\end{equation}
with $\mu = M_{\rm PBH}/\kappa M_{\rm H}$, $M_{\rm eq} \simeq 2.8 \times 10^{17} M_\odot$ the horizon mass at matter-radiation equality, $\Omega_{\rm DM} = 0.26$ the present-day DM density fraction~\cite{Planck:2018jri}, and
\begin{equation}
\label{eq:nl}
\Lambda = 1- \left( \delta_c(M_{\rm H}) - \frac{3}{2}\mu^{1/\gamma} \right)\,.
\end{equation}
The lower limit of the integral is set by the condition $\Lambda > 0$.
The total contribution of PBHs to today's energy budget is then given by
\begin{equation}
\label{eq:omega_bh}
    \Omega_{\rm PBH} = \Omega_{\rm DM} \int \mathcal{F}(M_{\rm PBH})\,{\rm d}\ln M_{\rm PBH}\,. 
\end{equation}
Eq.~\eqref{eq:fofm} clearly highlights the exponential dependence of the mass function on the collapse threshold $\delta_c(M_{\rm H})$, implying that even small variations in this parameter can lead to significant changes in the resulting PBH mass function. 
The collapse threshold depends on the equation of state of the Universe at the time of formation. This dependence is typically determined through numerical simulations, whose results are sensitive to the details of the collapse dynamics. Nevertheless, the threshold is found to increase monotonically with the EoS parameter in the range of $w$ considered here~\cite{Musco:2012au}.
As a consequence, when the EoS parameter $w = p/\rho$ drops below $1/3$, the corresponding decrease of the threshold $\delta_c$ enhances the PBH formation, leading to a resonant production of PBHs with masses comparable to the horizon mass at that time. For instance, in the case of degenerate MSSM masses (MSSM--1 and MSSM--3 in Fig.~\ref{fig:w}), the EoS softening results in a reduction of $\delta_c$ by about $5\%$, which in turn translates into an enhancement of the PBH abundance by more than one order of magnitude. These results also depend on the collapse dynamics and on the shape of the primordial perturbations~\cite{Musco:2023dak}, which is kept fixed in this analysis.
\begin{figure*}[t!]
    \centering
    \includegraphics[width=0.45\textwidth]{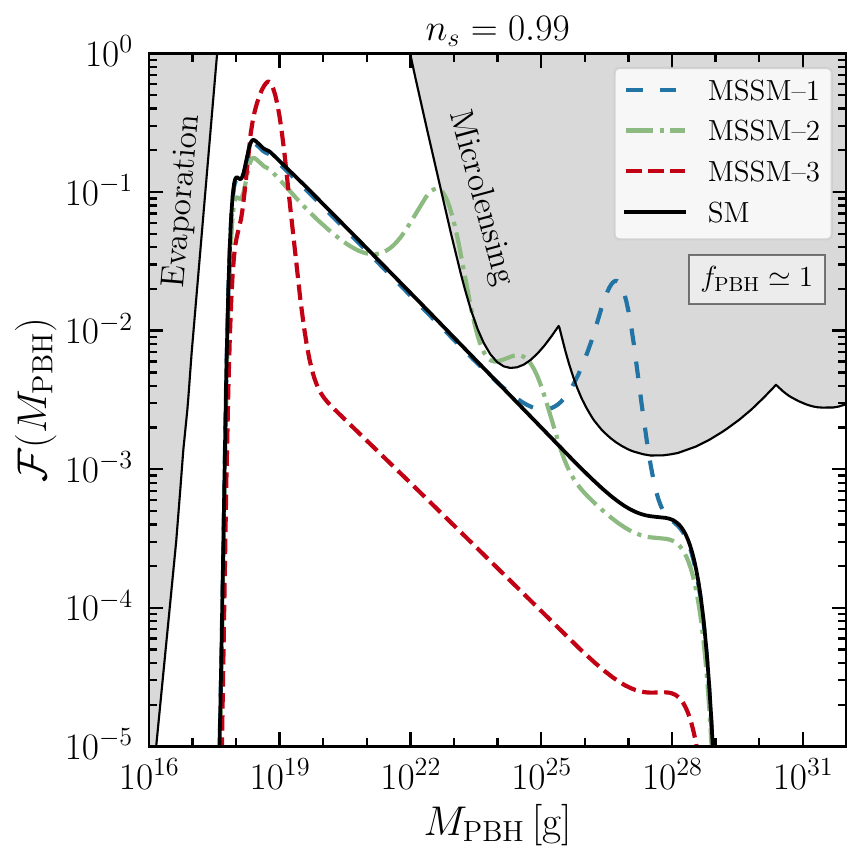}
    \hspace{0.05\textwidth}
    \includegraphics[width=0.45\textwidth]{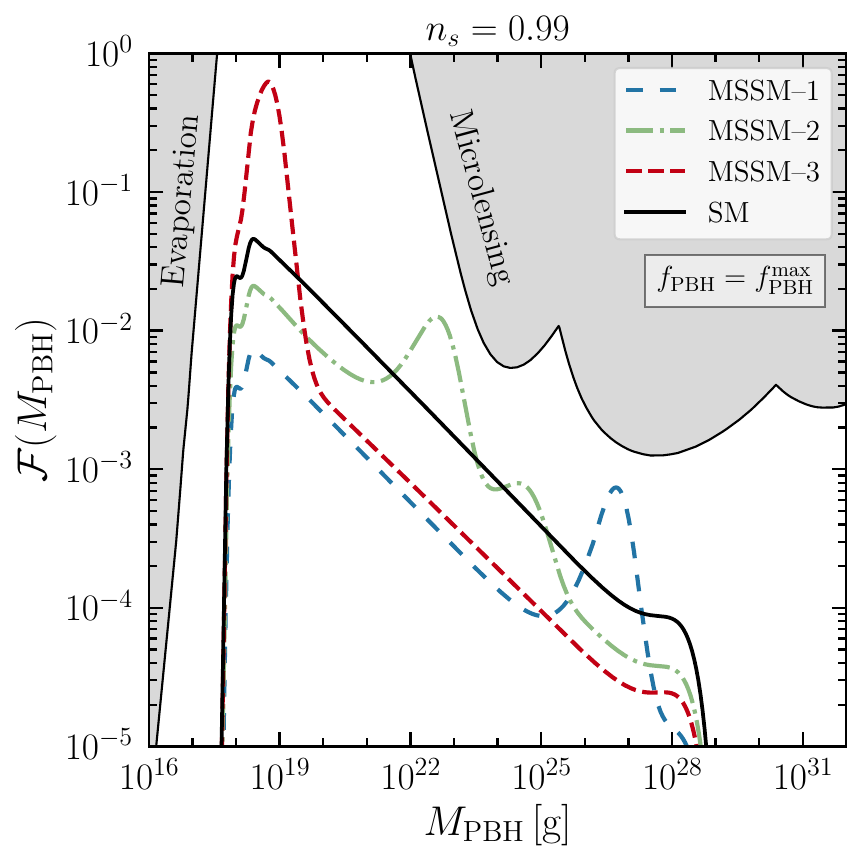}
        \caption{Primordial black hole mass fraction $\mathcal{F}(M_{\rm PBH})$ as a function of the PBH mass assuming $n_s = 0.99$. The colored curves correspond to the MSSM benchmark realizations listed in Sec.~\ref{sec:eos}, while the black curve shows the mass fraction computed assuming the SM equation of state.
        In the left panel, PBHs account for the total DM abundance, while in the right panel the mass functions satisfy $f_{\rm PBH} = f_{\rm PBH}^{\rm max}$ to be consistent with current observational constraints. Also displayed are the monochromatic evaporation \cite{Lehmann:2018ejc, Laha:2019ssq, Boudaud:2018hqb, Clark:2016nst, Dasgupta:2019cae, Laha:2020ivk, Coogan:2020tuf, Kim:2020ngi, Mittal:2021egv, Su:2024hrp, Tan:2024nbx, DelaTorreLuque:2024qms, Khan:2025kag} and microlensing \cite{Macho:2000nvd, EROS-2:2006ryy, Garcia-Bellido:2017imq, Oguri:2017ock, Croon:2020ouk, Blaineau:2022nhy, Leung:2022vcx, Esteban-Gutierrez:2023qcz, Mroz:2024wia}
        constraints, compiled using the repository~\cite{Kavanagh_PBHbounds}.}
    \label{fig:fpbh}
\end{figure*}

For our purposes, we need a primordial curvature power spectrum which is broad and nearly flat, so that our resulting mass function is sensible to different MSSM scales from $10^3$ to $10^7$~GeV. According to Eq.~\eqref{eq:temp}, these particle mass scales roughly correspond to a horizon mass $M_{\rm H} \sim 10^{26} ~\rm g$ and $M_{\rm H} \sim 10^{18}~ \rm g$, respectively. A power spectrum with these characteristics can be obtained in multi-field inflationary models~\cite{Ferrante:2023bgz, Stamou:2024lqf, Crescimbeni:2025ywm}.
We adopt a truncated power-law parametrization at the small scales relevant for PBH formation, defined as
\begin{equation}
\label{eq:ps}
P_\zeta(k)=
\begin{cases}
A_*\,\left(\dfrac{k}{k^*}\right)^{n_s-1} & k_1\leq k \leq k_2 \\[6pt]
0 & \text{otherwise}
\end{cases}\,,
\end{equation}
where $k_1 =  10^7 \, \mathrm{Mpc}^{-1}$ and $k_2 =  10^{13} \, \mathrm{Mpc}^{-1}$ (see Eq.~\eqref{eq:pivot}), while $n_s$ denotes the spectral index. For simplicity, we neglect the running of the spectral index, $\alpha_s = \mathrm{d}n_s/\mathrm{d}\ln k$, instead we vary $n_s$ from 0.980 to 0.995. The parameter $A_*$ is treated as a free normalization, setting the amplitude of the spectrum at the pivot scale $k^*$, corresponding to a characteristic PBH mass of $\mathcal{O}(10^{23}\mathrm{g})$.

The amplitude $A_*$ can be fixed by requiring that PBHs account for the total DM abundance, namely
\begin{equation}
f_{\rm PBH} \equiv \frac{\Omega_{\rm PBH}}{\Omega_{\rm DM}} =\int \mathcal{F}(M_{\rm PBH}) \, \mathrm{d}\ln M_{\rm PBH} \simeq 1 \,.
\end{equation}
Imposing this condition, we find $A_*\sim\mathcal{O}(10^{-3})$ for all the values of $n_s$ considered. This result indicates that the amplitude of the primordial curvature perturbations must be enhanced by several orders of magnitude with respect to its value at CMB scales \cite{Planck:2018jri}, to produce a sizeable amount of PBHs. We also find that the value of $A_*$ required to account for the total DM abundance is largely insensitive to $n_s$, due to the exponential dependence of the PBH abundance on the variance of the density fluctuations.

At this stage, we must  compare our results with various constraints on PBHs as DM candidates available in the literature~\cite{Green:2020jor, Carr:2026hot}. These bounds arise from different observational probes, depending on the PBH mass, and are typically derived under the assumption of a monochromatic PBH mass function. To extend them to a general mass distribution, we follow the prescription of Ref.~\cite{Carr:2017jsz} and define the maximum observationally allowed PBH fraction in DM as
\begin{equation}
\label{eq:fmax}
f_{\rm PBH}^{\rm max} = \left[ \sum_i \left( \int \frac{\mathcal{F}(M_{\rm PBH})}{\mathcal{F}^{\rm max}_i(M_{\rm PBH})} 
\frac{{\rm d}M_{\rm PBH}}{M_{\rm PBH}} \right)^2 \right]^{-1/2}\,,
\end{equation}
where $\mathcal{F}^{\rm max}_i(M_{\rm PBH})$ denotes the maximum PBH dark matter fraction allowed by monochromatic constraints for the $i$-th observable. The quantity $f_{\rm PBH}^{\rm max}$ directly quantifies the degree to which a given extended mass function is compatible with the observational bounds: values $f_{\rm PBH}^{\rm max} \geq 1$ indicate that PBHs can account for all dark matter, while $f_{\rm PBH}^{\rm max} < 1$ indicate that PBHs can contribute at most a fraction $f_{\rm PBH}^{\rm max}$ of the total DM abundance.
This procedure is valid when PBHs of different masses contribute approximately independently to a given observable, so that the total signal can be expressed as a linear superposition over the mass function.  This condition is well satisfied for microlensing constraints which are the most relevant for our analysis and, to a good approximation, for evaporation bounds.

In Fig.~\ref{fig:fpbh} we show the PBH mass functions corresponding to the benchmark scenarios introduced in Sec.~\ref{sec:eos}. The left panel corresponds to the case where we assume $f_{\rm PBH} = 1$, with the mass functions for the MSSM scenarios peaking at $M_{\rm PBH} \sim M_{\rm H}(T=m_i)$ with $i=b,f,h$, as expected.
The right panel shows the same mass functions rescaled by $f_{\rm PBH} = \min \{ f_{\rm PBH}^{\rm max}, 1 \}$, where the minimum ensures that unphysical cases with $f_{\rm PBH} > 1$ are excluded. 
For the displayed cases, PBHs in the SM, MSSM--1, and MSSM--2 scenarios cannot account for the entirety of the dark matter, with their relative abundance constrained to be below $\sim 10\%$, and the most stringent bound applying to MSSM--1, which is limited to $\sim 5\%$ of the total DM abundance. The only exception is the MSSM--3 mass function, which features a narrow peak at $M_{\rm PBH} \sim 10^{19}\,\mathrm{g}$ and remains consistent with current observational constraints.

%====================================
\section{Results}
\label{sec:res}
%====================================

We compute the equation of state by varying the three masse scales $(m_f, m_b, m_h)$ from $10^3 \rm GeV$ to $10^7 \rm GeV$ in order to assess the impact on PBH formation due to the variations of the EoS parameter in the MSSM. For each mass configuration, we first evaluate the resulting PBH mass function by fixing the spectral index $n_s$ and adjusting the amplitude of the primordial power spectrum so as to obtain $f_{\rm PBH} \simeq 1$. We quantify the enhancement in PBH production through the ratio $\mathcal{F}_{\rm MSSM}(M_{\rm PBH})/\mathcal{F}_{\rm SM}(M_{\rm PBH})$, where $\mathcal{F}_{\rm SM}(M_{\rm PBH})$ is obtained using the SM equation of state.
\begin{figure}[t!]
    \centering
    \includegraphics[width=0.47\textwidth]{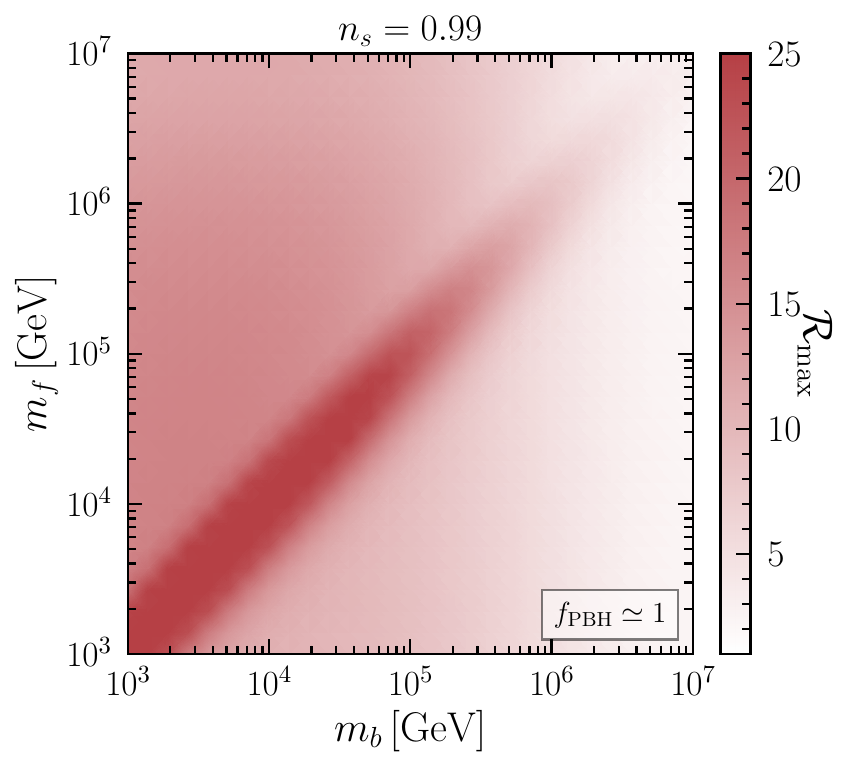}
       \caption{Maximum enhancement of the PBH mass fraction in the MSSM relative to the SM case, $\mathcal{R}_{\rm max}$ (see Eq.~\eqref{eq:ratio}, shown in the $(m_f, m_b)$ plane for fixed $m_h = 10^3 \, \rm GeV$ and $n_s = 0.99$. For each scenario, we here consider $f_{\rm PBH} \simeq 1$.}
    \label{fig:ratio}
\end{figure}

In Fig.~\ref{fig:ratio} we show the maximum value 
\begin{equation}
    \label{eq:ratio}
    \mathcal{R}_{\rm max} \equiv \max_{M_{\rm PBH}} \left[\frac{\mathcal{F}_{\rm MSSM}(M_{\rm PBH})}{\mathcal{F}_{\rm SM}(M_{\rm PBH})}\right]
\end{equation}
in the $(m_f, m_b)$ plane, fixing $m_h = 10^3 \,\rm GeV$. This choice is motivated by the relatively small number of degrees of freedom associated with the heavy Higgs, which leads to a milder modification of the EoS compared to the fermionic and bosonic superpartners. A clear enhancement of the PBH abundance appears along the $m_f = m_b$ direction. This behavior is expected as the drop in the EoS is maximized when multiple species decouple simultaneously, {\it i.e.} for nearly degenerate masses. We find the maximum value for the ratio to be $\mathcal{R}_{\rm max} \simeq 40$ for $m_b = m_f\simeq 3.7 \times 10^3 \, \rm GeV$ and $n_s = 0.995$. 
A complementary feature of our results is that the peak of the PBH mass function is largely set by $m_b$, reflecting the dominant contribution of bosonic degrees of freedom. In most cases, the peak mass is well approximated by
\begin{equation}
\label{eq:mpeak}
    M_{\rm PBH}^{\rm peak} \simeq 5 \times 10^{22} \left(\frac{10^5~{\rm GeV}}{m_b}\right)^2 \rm g \, ,
\end{equation} 
with only a weak dependence on $m_f$ and $m_h$.
\begin{figure}[t!]
    \centering
    \includegraphics[width=0.48\textwidth]{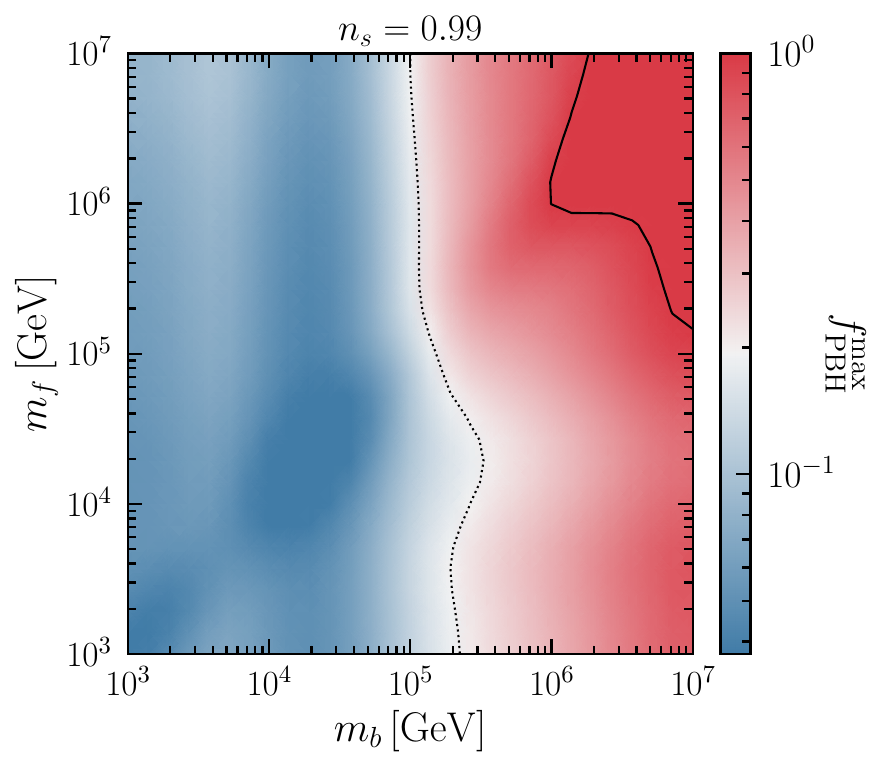}
    \caption{Maximum allowed fraction of DM in the form of PBHs, $f_{\rm PBH}^{\rm max}$, in the $(m_b, m_f)$ parameter space with $m_h = 10^3~{\rm GeV}$ and $n_s=0.99$. The color scale is chosen such that the white region marks where the SM and MSSM scenarios predict the same value of $f_{\rm PBH}^{\rm max}$, {\it i.e.} where the effect of observational constraints is unchanged. The dotted black curve further outlines this region, while the solid contour encloses the parameter space in which PBHs can account for all of the DM abundance. To the right (left) of the dotted curve, the allowed DM fraction in PBHs is larger (smaller) than in the SM case.}
    \label{fig:fDM}
\end{figure}

We then confront the resulting mass functions with existing observational constraints in Fig.~\ref{fig:fDM}, where for each scenario we show the quantity $f_{\rm PBH}^{\rm max}$ given in Eq.~\eqref{eq:fmax}. In the SM case, PBHs cannot account for the total DM abundance for any of the $n_s$ values considered, as the corresponding mass functions are excluded by evaporation and microlensing bounds. The white region in the plot denotes the subset of parameter space in which the MSSM and SM scenarios yield identical values of $f_{\rm PBH}^{\rm max}$.

Several MSSM realizations shift and enhance the peaks of the mass function toward regions of parameter space that are currently weakly constrained, the asteroid-mass window. Remarkably, we find that, for specific combinations of $(n_s, m_f, m_b, m_h)$, the resulting mass function is compatible with primordial black holes accounting for the entirety of dark matter. For all the scenarios considered, this improvement in compatibility with monochromatic constraints occurs for $m_b \gtrsim  10^5 \,\rm GeV$, while for smaller values the resulting mass functions are more strongly excluded than in the SM case. According to Eq.~\eqref{eq:mpeak}, $m_b \gtrsim 10^5 \,\rm GeV$ corresponds to a peak mass $M^{\rm peak}_{\rm PBH} \lesssim  10^{22}\,\rm g$, which lies within the asteroid-mass window where monochromatic constraints are comparatively weaker or absent. 

\begin{figure}[t!]
    \centering
    \includegraphics[width=0.95\linewidth]{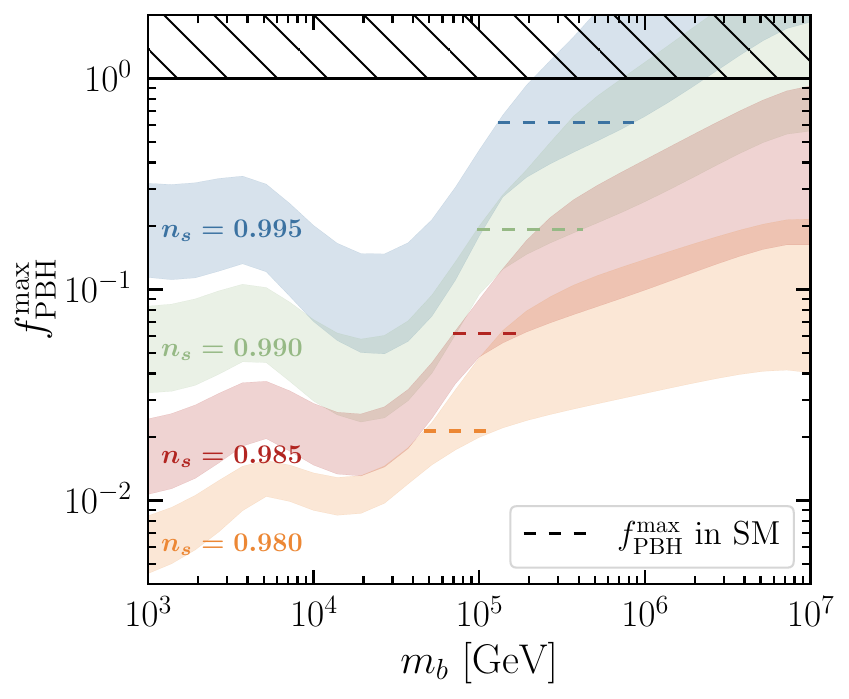}
    \caption{Maximum allowed DM fraction in PBHs, $f_{\rm PBH}^{\rm max}$, as a function of $m_b$ for different values of the spectral index $n_s$. Shaded bands are obtained by varying $m_f$ and $m_h$, with their edges corresponding to the minimum and maximum values of $f_{\rm PBH}^{\rm max}$. Dashed lines show the SM value of $f_{\rm PBH}^{\rm max}$.}
    \label{fig:fns}
\end{figure}

In Fig.~\ref{fig:fns} we present the maximum allowed PBH dark matter fraction, $f_{\rm PBH}^{\rm max}$, as a function of $m_b$ for different values of the spectral index $n_s$. For each $n_s$, the shaded bands illustrate the range of the allowed values obtained by varying $m_f$ and $m_h$ over the parameter space considered. The extent of these bands highlights the sensitivity of the constraints on the PBH abundance to the underlying mass spectrum, with larger widths indicating a stronger dependence on the fermionic and Higgs mass scales. The horizontal dashed segments indicate the corresponding $f_{\rm PBH}^{\rm max}$ in the SM scenario. The features induced by the MSSM equation of state in the PBH mass function are largely insensitive to the spectral index, as the varying $n_s$ primarily produces an overall vertical shift of the curves.

We find that, for our choice of parameters, a nearly scale-invariant power spectrum leads to a mass function that is in better overall agreement with current observational constraints. Indeed, only for spectral index values close to unity ($n_s = 0.990$ and $n_s = 0.995$ in our analysis), the features induced by the MSSM equation of state significantly improve the compatibility with monochromatic bounds, allowing PBHs to constitute the entirety of dark matter. For smaller spectral index values, the mass function develops a high-mass tail that is constrained by the absence of microlensing events in observations.

%====================================
\section{Conclusions}
\label{sec:conc}
%====================================

In this work, we have investigated a scenario in which a heavy supersymmetric sector becomes non-relativistic at temperatures $T \gtrsim 1\,\mathrm{TeV}$, inducing a temporary softening of the equation of state and enhancing the PBH production in the asteroid-mass range. We computed the equation-of-state parameter for different MSSM realizations, parametrized by the masses of the fermionic, bosonic, and heavy Higgs sectors, which we assumed to be degenerate within each sector for simplicity. We then derive the resulting PBH mass functions within the critical collapse framework, adopting a minimal phenomenological parametrization of the primordial curvature power spectrum. We quantified the enhancement in PBH formation by comparing the MSSM and SM predictions, finding that the dominant peaks in the mass function are associated with the mass scale of the bosonic superpartners.

We showed that, for all the spectral index values considered, PBHs cannot account for the entirety of dark matter in the SM scenario, as the corresponding mass functions are excluded by current monochromatic constraints. In contrast, we found that specific MSSM equations of state can both enhance and shift the PBH mass function toward regions of parameter space that remain weakly constrained, in particular within the asteroid-mass window. For suitable combinations of supersymmetric mass scales, PBHs can consistently account for the full DM abundance. Interestingly, this occurs in regions of parameter space where the relative enhancement of the mass function is moderate, rather than maximal.

Overall, our results demonstrate that modifications of the high-energy particle content above the electroweak scale can leave observable imprints on small-scale structure formation through their impact on PBH production. A more complete assessment would require embedding the primordial power spectrum within a concrete inflationary framework, which we leave for future work.

Finally, we emphasize that our analysis have implications for particle DM searches, as supersymmetric models naturally predict additional DM candidates that freeze out and contribute to the relic abundance. In scenarios where PBHs account for the entirety of the dark matter, the contribution from particle DM must necessarily be negligible or absent, implying strong constraints on the underlying particle physics framework. The interplay between PBH and particle DM components has not been addressed here and will be the subject of future investigations.

\section*{Acknowledgments}

AB thanks A. J. Iovino and A. Caravano for insightful discussions on PBH formation and inflation.
We acknowledge the support by the research project TAsP (Theoretical Astroparticle Physics) funded by the Istituto Nazionale di Fisica Nucleare (INFN). 

\bibliography{bibliography}
\end{document}